\documentclass[aps,prl,twocolumn,superscriptaddress]{revtex4}

\usepackage{graphicx}
\usepackage{amsmath}
\usepackage{amssymb}
\usepackage{braket}

\begin{document}

\title{Competition between Chaotic and Non-Chaotic Phases in a Quadratically Coupled Sachdev-Ye-Kitaev Model}
\author{Xin Chen}
\thanks{They contribute equally to this work. }
\affiliation{Institute for Advanced Study, Tsinghua University, Beijing, 100084, China}
\author{Ruihua Fan}
\thanks{They contribute equally to this work. }
\affiliation{Institute for Advanced Study, Tsinghua University, Beijing, 100084, China}
\affiliation{Department of Physics, Peking University, Beijing, 100871, China }
\author{Yiming Chen}
\thanks{They contribute equally to this work. }
\affiliation{Institute for Advanced Study, Tsinghua University, Beijing, 100084, China}
\affiliation{Department of Physics, Tsinghua University, Beijing, 100084, China }

\author{Hui Zhai}
\thanks{Electronic address: hzhai@tsinghua.edu.cn }
\affiliation{Institute for Advanced Study, Tsinghua University, Beijing, 100084, China}
\affiliation{Collaborative Innovation Center of Quantum Matter, Beijing, 100084, China}

\author{Pengfei Zhang}
\thanks{Electronic address: PengfeiZhang.physics@gmail.com }
\affiliation{Institute for Advanced Study, Tsinghua University, Beijing, 100084, China}

\date{\today}

\date{\today }

\begin{abstract}
The Sachdev-Ye-Kitaev (SYK) model is a concrete solvable model to study non-Fermi liquid properties, holographic duality and maximally chaotic behavior. In this work, we consider a generalization of the SYK model that contains two SYK models with different number of Majorana modes coupled by quadratic terms. This model is also solvable, and the solution shows a zero-temperature quantum phase transition between two non-Fermi liquid chaotic phases. This phase transition is driven by tuning the ratio of two mode numbers, and a Fermi liquid non-chaotic phase sits at the critical point with equal mode number. At finite temperature, the Fermi liquid phase expands to a finite regime. More intriguingly, a different non-Fermi liquid phase emerges at finite temperature. We characterize the phase diagram in term of the spectral function, the Lyapunov exponent and the entropy. Our results illustrate a concrete example of quantum phase transition and critical regime between two non-Fermi liquid phases.
\end{abstract}

\maketitle

\textit{Introduction.} The Landau's Fermi liquid is a very fundamental concept in physics that describes a large variety of interacting fermion models \cite{FL}. Only until recent years, some strongly correlated materials are discovered where a Fermi liquid description fails \cite{NF1}. However, due to the strong interaction in these materials, theoretical investigations of the non-Fermi liquid with controlled approximations are quite limited, which makes a solvable model exhibiting non-Fermi liquid behavior very valuable. Recently, a model named the Sachdev-Ye-Kitaev (SYK) model, describing $N$ Majorana fermions with all-to-all random interaction, has been proposed \cite{Kitaev1, Kitaev2,SY,Comments}. In the large-$N$ limit, this model is exactly solvable and shows non-Fermi liquid behavior. That is one of the reasons that the SYK model draws lots of attentions recently \cite{spectrum1,spectrum2,spectrum3,Liouville,Liouville2,bulk Yang,bulk spectrum Polchinski,bulk2,bulk3,bulk4,bulk5,syk-bh,SYK-new scramble,SYK-new bh}. Various extensions of this model \cite{numerics wenbo,wenbo susy,susy2,Yingfei1,Yingfei2,Altman,sk jian,yyz condensation,generalization 1,thermal transport,high-D1,high-D2-con,generalization 2,transition1,gen-new 1,no disorder1,no disorder2,no disorder3,no disorder4,no disorder5,no disorder6,no disorder7,no disorder8,no disorder9,no disorder10} have also been studied to illustrate its non-Fermi liquid properties.  

To distinguish a non-Fermi liquid from a Fermi liquid, the dynamical properties have also been highlighted in recent studies, apart from their difference in the spectrum function \cite{subir1, subir2,Ashcroft}. Let's consider the local thermalization time $\tau$. For a Fermi liquid, generally $\tau$ is proportional to $1/T^2$. While for a non-Fermi liquid, it is widely believed that $\tau \sim  \hbar/(k_\text{B}T)$. Moreover, recent studies also reveal that $\tau$ is closely related to the Lyapunov exponent $\lambda$ defined from the out-of-time-ordered correlation function (OTOC) \cite{Kitaev1}. It has been shown that the Lyapunov exponent of a quantum system is bounded by $2\pi k_\text{B}T$ \cite{prove}. For a Fermi liquid, $\lambda$ usually behaves as $\sim T^2$ at low temperature \cite{Altman,OTOC-Keldysh}, and is much smaller than the bound; while if a quantum system is holographically dual to a gravity system, it is considered to be maximally chaotic and $\lambda$ should saturate the bound \cite{bh1,bh2,bh3}. Such a holographic quantum system is normally a non-Fermi liquid. There are strong evidences that the SYK model displays a duality to AdS$_2$ gravity with a black hole \cite{bulk Yang,bulk spectrum Polchinski}. The two- and four-point correlation functions of the SYK model can be explicitly calculated exactly and its $\lambda$ indeed saturates the bound at the low-energy limit. That is another reason why the SYK model is so interesting.   

In this work we will consider a natural generalization of the SYK model, that is, two SYK models coupled by a quadratic coupling term. This model is also solvable in the large-$N$ limit. We will show that at zero temperature, there exist two maximally chaotic non-Fermi liquid phases separated by a non-chaotic Fermi liquid point. More interestingly, at finite temperature, a different non-Fermi liquid phase emerges in the quantum critical regime. We hope that this concrete example will shed light on the understanding of quantum phase transition between non-Fermi liquid phases. 

\begin{figure}[tp]
\includegraphics[width=2.2 in]
{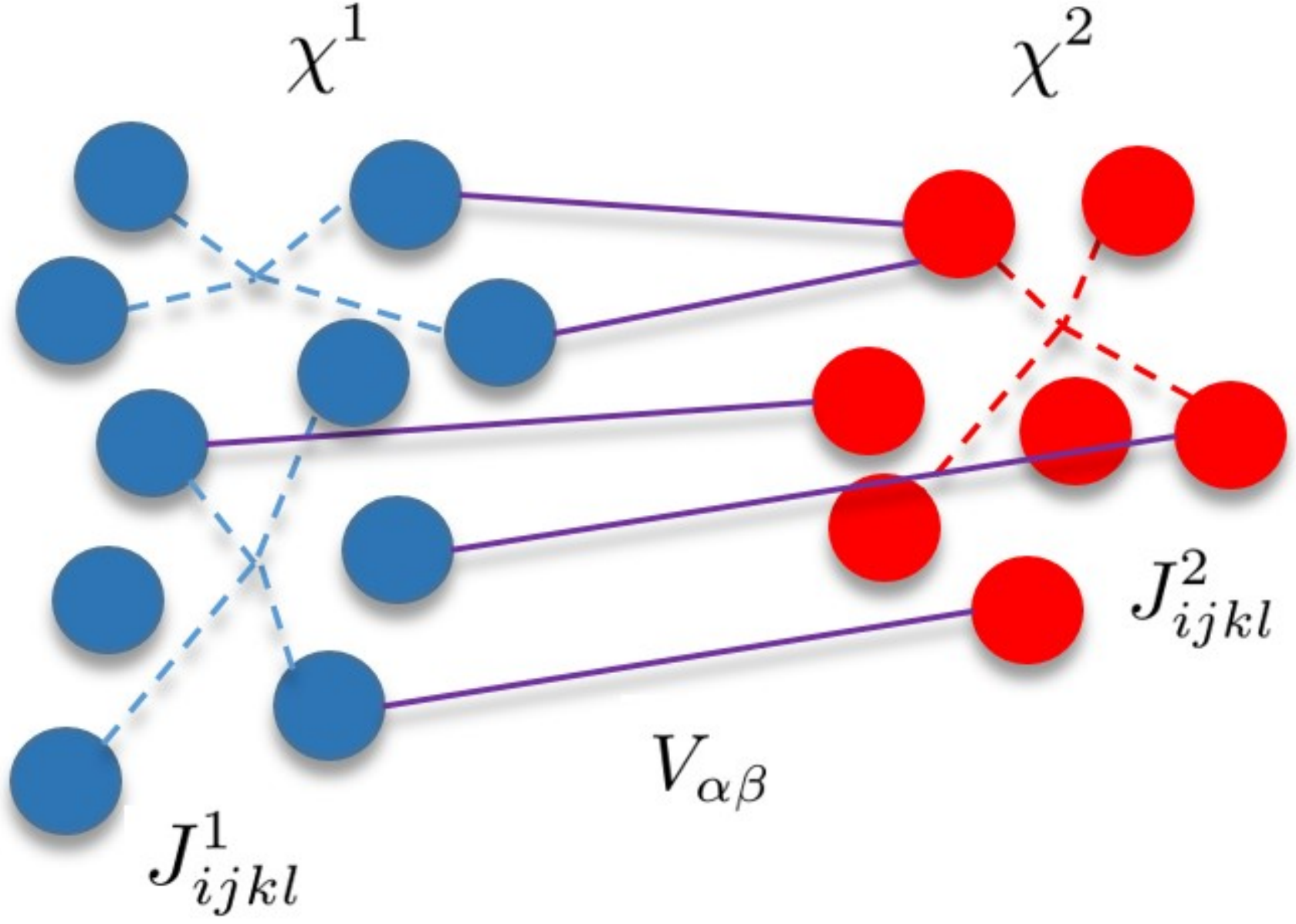}
\caption{Schematic of the model that two SYK$_4$ models with different numbers of modes are coupled by quadratic couplings.  }
\label{model}
\end{figure}

\textit{Model and Phases.} The system we considered is schematically illustrated in Fig. \ref{model}, and its Hamiltonian is given by
\begin{align}
&\hat{H}=\sum\limits_{\xi=1,2}\hat{H}^\xi_{SYK}+i \sum_{\alpha=1}^{N_1}\sum\limits_{\beta=1}^{N_2}V_{\alpha\beta} \chi^1_\alpha \chi^2_\beta,\\
&\hat{H}^\xi_{SYK}=\frac{1}{4!}\sum_{ijkl=1}^{N_\xi}J^\xi_{ijkl}\chi^\xi_i\chi^\xi_j\chi^\xi_k\chi^\xi_l,
\end{align}
where each $\hat{H}^\xi_{SYK}$ is a SYK$_4$ model with Majorana fermions $\chi^\xi_i$ \cite{footnote}, and for the $\xi$th SYK model, there are totally $N_\xi$ modes. $J^\xi_{ijkl}$ and $V_{\alpha\beta}$ are all random with zero expectation value and 
\begin{align}
\overline{(J^{\xi}_{ijkl})^2} = \frac{3!J^2}{N_\xi^3}, \quad \overline{V^2_{\alpha\beta}} = \frac{V^2}{\sqrt{N_1N_2}},
\end{align}
up to some permutation of indexes. In the large-$N$ limit, both two $N_\xi\rightarrow\infty$ and $p=N_2/N_1$ is fixed. The coupling is a SYK$_2$ type term.

In this model, there are three independent parameters chosen as $p$, $V/J$ and $1/(\beta\sqrt{J^2+V^2})$ ($\beta=1/(k_\text{B}T)$). In this work we will discuss the transition or crossover between different phases in terms of these three parameters. Let us first analyse possible phases of this model. Apart from a free fermion phase, other phases are denoted by the notation $(a_1,a_2)$, where $\alpha_\xi$ denotes the scaling dimension of operator $\chi^\xi$.

(1) $(1/4,3/4)$ phase. Because $[\chi^1]=1/4$ and $[\chi^2]=3/4$, the first SYK$_4$ term and the SYK$_2$ term are marginal. Note that the model is symmetric by exchanging index $1\leftrightarrow 2$ and $p\rightarrow 1/p$, under which the $(1/4,3/4)$ phase becomes $(3/4,1/4)$. 

(2) $(1/2,1/2)$ phase. Both two $[\chi^\xi]=1/2$, and only the SYK$_2$ term is marginal. 

(3) $(1/4,1/4)$ phase. Because two $[\chi^\xi]=1/4$, both two SYK$_4$ terms are marginal, but the SYK$_2$ term is relevant, thus this phase is not a stable phase at zero temperature except for $V=0$. 

\textit{Green's Function.} With the standard large-$N$ method for the SYK model, we obtain coupled self-consistent equations for the imaginary-time Green's function and the self-energy as
\begin{align}
&\Sigma_1(\tau) = J^2G_1^3(\tau) + V^2\sqrt{p}G_2(\tau),\label{two-point1}\\
&\Sigma_2(\tau) = J^2G_2^3(\tau) + V^2\sqrt{\frac{1}{p}}G_1(\tau). \label{two-point2}
\end{align}
where we define the time-ordered two-point Green's function as $G_\xi(\tau) \delta_{ij}= \braket{\mathcal{T}\chi^\xi_i(\tau)\chi^\xi_j(0)}_\beta$, the self-energy $\Sigma_\xi(\omega_n)= -i\omega_n - G_\xi^{-1}(\omega_n)$, and only the diagonal terms of the Green's function enters because of the disorder average. Here $\tau$ satisfies the periodic boundary condition between zero and $\beta$. 

There are two ways we can proceed from here. First, we can consider a zero-temperature low-energy limit by taking $\beta\rightarrow \infty$. In this case, for different phases listed above, we drop the irrelevant terms in Eq.\ref{two-point1} and \ref{two-point2} when solving the equations. Then, the theory displays an emergent conformal symmetry. By using a proper ansatz obeying the symmetry constraint, the Green's function $G_\xi(\tau)$ can be solved in the imaginary time. This will be discussed later in detail. Assuming this conformal symmetry still holds at finite but low-temperature, and utilizing a conformal mapping of $\tau\rightarrow \tan (\pi \tau/\beta)$, the Green's function $G(\tau)$ at finite temperature can be obtained. Furthermore, by analytical continuation, one can obtain the retarded Green's function in real time and finite temperature, as well as its Fourier transform $G_{R,\xi}(\omega)$, from which we can determine the spectral function $A_\xi(\omega)=-(1/\pi)\text{Im} G_{R,\xi}(\omega)$. In this way, we can determine the characteristic features of different phases from the spectral functions. 

Second, by directly applying the analytical continuation to the self-consistent equations of Eq. \ref{two-point1} and \ref{two-point2}, we can obtain the self-consistent equation for the retarded Green's function $G_{R,\xi}(\omega)$.
By solving these equations directly with numerics, the spectrum function $A_\xi$ can also be calculated at finite temperature. This solution goes beyond the conformal limit. This will also be discussed later when we talk about the numerical results. 

The Lyapunov exponent is extracted from the out-of-time-ordered correlation function defined as \cite{Kitaev1,prove}
\begin{align}
F_{\xi\xi}(t) &= \frac{1}{N_\xi^2}\sum_{ij}\text{Tr}\left[y\chi^\xi_i(t)y\chi^\xi_j(0)y\chi^\xi_i(t)y\chi^\xi_j(0)\right],\\
F_{\xi^\prime \xi}(t) &= \frac{1}{N_\xi N_{\xi^\prime}}\sum_{i j} \text{Tr}\left[y\chi^{\xi^\prime}_i(t) y\chi^\xi_j(0) y\chi^{\xi^\prime}_i(t) y\chi^\xi_j(0)\right],
\end{align}
where $y=\exp(-\beta \hat{H}/4)$ and $\xi\neq \xi^\prime$. Using the Keldysh contour one can obtain that the disconnected part of $F$ follows a Bethe-Salpeter equation, and the exponential increasing of $F(t)\sim \exp(\lambda t)$ defines a Lyapunov exponent. Since $F_{\xi\xi}$ and $F_{\xi^\prime \xi}$ are coupled in the Bethe-Salpeter equation, they will give the same Lyapunov exponent. 

\textit{Analysis in the Conformal Limit.} In the conformal limit, results such as the spectral functions, the Lyapunov exponent and the entropy can be obtained analytically. Below we will first list the results in this limit.  

\textbf{A: The Spectral Function.} For the $(1/4,3/4)$ phase, the general ansatz for the Green's function will be $G_1(\tau)\propto \text{sgn}(\tau)/|\tau|^{1/2}$, and $G_2(\tau)\propto \text{sgn}(\tau)/|\tau|^{3/2}$. Dropping the irrelevant terms $J^2 G_2^3$ and $-i\omega_n$ in Eq. \ref{two-point1} and \ref{two-point2}, we can obtain the solution for the entire regime with $p<1$ and $V\neq 0$ \cite{Altman}. This gives rise to a divergent spectral function in $A_1(\omega)$ at low-energy at zero temperature, revealing a non-Fermi liquid behavior. Similarly, the ansatz for $(3/4,1/4)$ phase can be found for the entire regime $p>1$ and $V\neq 0$, and the same non-Fermi liquid behavior can be found in $A_2$. 

For the $(1/2,1/2)$ phase, we take the ansatz $G_\xi(\tau)\propto \text{sgn}(\tau)/|\tau|$ for both $\xi=1,2$. Dropping the irrelevant terms of both $J^2 G_\xi^3$ for $\xi=1,2$ and $-i\omega_n$ in Eq. \ref{two-point1} and \ref{two-point2}, the solution can only be found when $p=1$ and $V\neq 0$. 
Its corresponding spectral function at zero frequency is finite at zero temperature, and therefore it is a Fermi liquid phase. Hence, the analysis above shows that, at zero-temperature with a fixed finite $V/J$, there will be a transition between two non-Fermi liquid phases ($(1/4,3/4)$ at $p<1$ and $(3/4,1/4)$ at $p>1$) with a Fermi liquid phase ($(1/2,1/2)$) sitting at the critical point ($p=1$).

In the zero-temperature limit, the $(1/4,1/4)$ phase only exists when $V=0$. In this case, the model becomes two decoupled SYK$_4$ models, and the solution is $G_\xi(\tau)\propto \text{sgn}(\tau)/|\tau|^{1/2}$ and the spectral function exhibits non-Fermi liquid behavior for both $\xi=1,2$ \cite{Comments}. 

\begin{figure}[tp]
\includegraphics[width=3.2 in]
{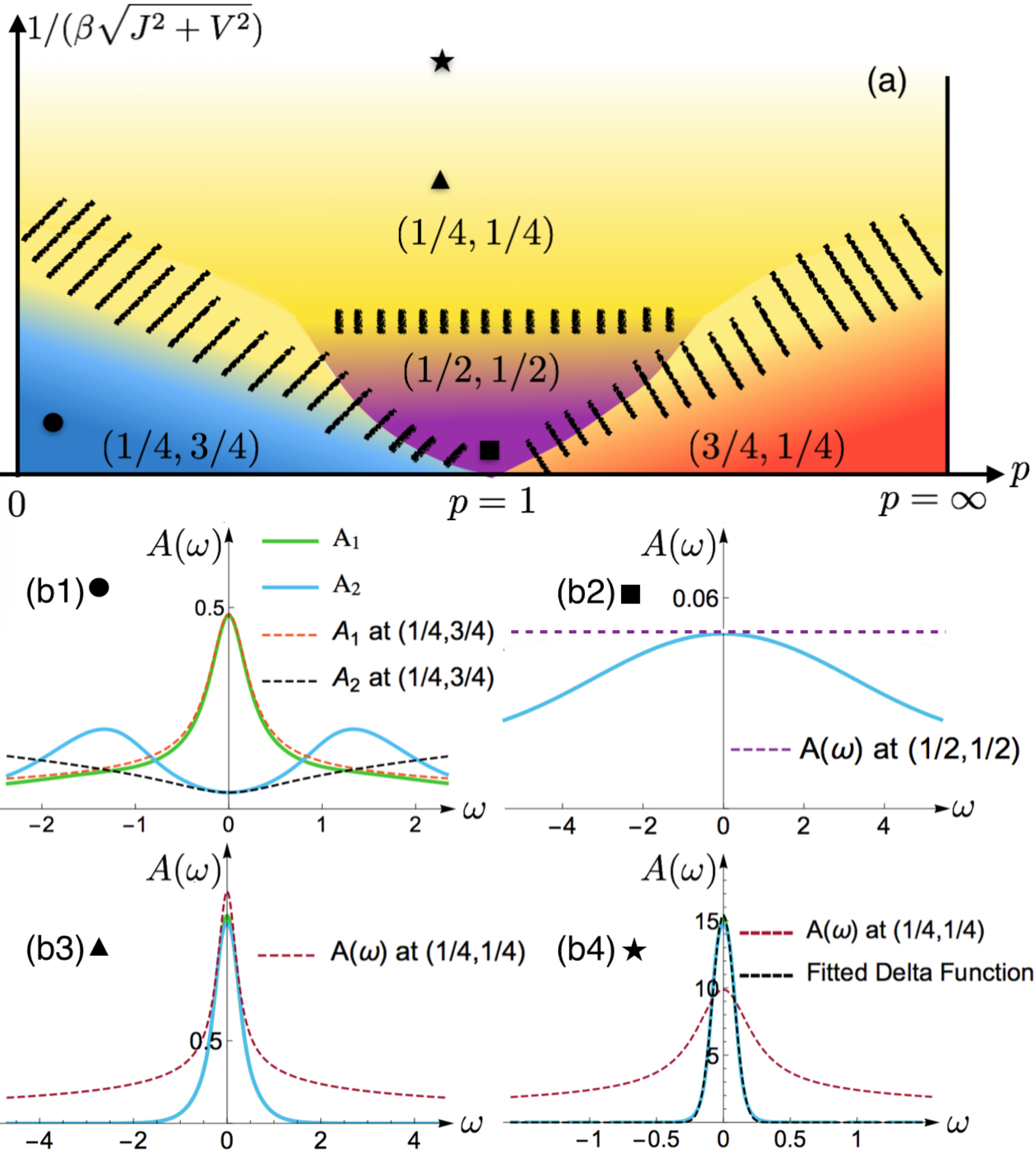}
\caption{(a) Schematic of a finite temperature phase diagram in terms of $p=N_2/N_1$ and temperature $k_\text{B}T$ in unit of $\sqrt{J^2+V^2}$, for a fixed $V/J=0.2$. At zero temperature, two non-Fermi liquid phases $(1/4,3/4)$ and $(3/4,1/4)$ are separated by a Fermi liquid point $(1/2,1/2)$ at $p=1$. At finite temperature, the color scheme and the dashed lines indicate the crossover between different phases. (b1-b4) The spectral functions $A(\omega)$ at four different representing points in the phase diagram as marked in (a). $A$ is in unit of $\beta$ and $\omega$ is in unit of $1/\beta$. The dashed lines (except for the fitted $\delta$-function) are results from the conformal limit analysis; while the solid lines are obtained from numerical solutions of the real time retarded Green's functions. In (b2-b4), two solid lines nearly coincide with each other and their difference is hard to see. (b1-b4) are computed at $\{p, 1/(\beta\sqrt{J^2+V^2})\}=\{0.1,0.2\}$, $\{1,0.005\}$, $\{0.9, 0.2\}$ and $\{0.9,0.9\}$, respectively.  }
\label{spectral}
\end{figure}

\textbf{B: The Lyapunov Exponent.} Following the standard procedure of solving the SYK model in the conformal limit\cite{Comments}, in our case we find all non-Fermi liquid $(1/4,3/4)$, $(3/4,1/4)$ and $(1/4,1/4)$ phases are maximally chaotic and display a Lyapunov exponent of $2\pi/\beta$ \cite{Altman}; while the Fermi liquid phase $(1/2,1/2)$ is not chaotic. 

\textbf{C: The Entropy.} With the solution for the two-point Green's functions, the free-energy of the system can also be obtained in the large-$N$ limit, with which the zero-temperature entropy can be calculated. It is also straightforward to show that for both $(1/4,3/4)$ and $(3/4,1/4)$ phases, the entropy normalized as $S/N_1$ is $|1-p|S_{SYK}$, where $S_{SYK}\sim 0.2324$ is the entropy for a single SYK model \cite{Comments}; while for the $(1/4,1/4)$ phase, the entropy will be $(1+p)S_{SYK}$. The difference in entropy can be used to distinguish $(1/4,1/4)$ and $(1/4,3/4)$ (or $(3/4,1/4)$) phase. The entropy of the Fermi liquid $(1/2,1/2)$ phase at $p=1$ vanishes at zero-temperature.

\textit{Phase Diagram.} Fig. \ref{spectral}(a) is the central result of this paper. This is a phase diagram in terms of temperature and $p=N_2/N_1$, with a fixed small $V/J$. At zero temperature, as discussed above, two non-Fermi liquid phases are separated by a Fermi liquid point; and at finite temperature, this point expands into a finite regime around the critical point. As temperature increases, a different phase $(1/4,1/4)$ emerges. This is very interesting for at least two reasons. First, as we discussed above, this phase does not exist at zero-temperature for any finite $V/J$, and it emerges only at finite temperature and in the quantum critical regime. Secondly, when $p\neq 1$ the model is not symmetric under exchanging index $1$ with $2$, while this phase does. It means that there is a kind of emergent $Z_2$ symmetry. 

At finite temperature there are no sharp boundaries between these phases. To roughly outline each regime, we first compute the spectral function directly from numerically solving the self-consistent equations for the real time retarded Green's function, as shown by the solid lines in Fig. \ref{spectral}(b1-b4). Then we compare the spectral functions to the characteristic features of the spectral functions for aforementioned different phases in the low-energy conformal limit, as shown by the dashed lines of Fig. \ref{spectral}(b1-b4). Due to the symmetry, we only show four representing points in the regime with $p<1$. Fig. \ref{spectral}(b1) shows that $A_1(\omega)$ displays a peak and $A_2(\omega)$ displays a dip at low-energy, and the low-energy behaviors of both $A_\xi(\omega)$ ($\xi=1,2$) are consistent with the spectral function of $(1/4,3/4)$ phase obtained in the conformal limit. In Fig. \ref{spectral}(b2), $A_1(\omega)$ and $A_2(\omega)$ coincide with each other and they are both much rounder. In fact their low-energy limits are consistent with results from the conformal limit of the $(1/2,1/2)$ phase. As temperature increases, in Fig. \ref{spectral}(b3), $A_1(\omega)$ and $A_2(\omega)$ still coincide with each other, despite that we already choose $p \neq 1$ to slightly derivate from the $p=1$ critical point. In contrast to the case of (b2), their low-energy behavior displays a rather sharp peak and is consistent with the conformal limit results of the $(1/4,1/4)$ phase. This is one evidence for the emergent $(1/4,1/4)$ phase. When further increasing the temperature, Fig. \ref{spectral}(b4) shows that the peak structure of $A_\xi(\omega)$ ($\xi=1,2$) deviates from the $(1/4,1/4)$ behavior and becomes more consistent with a $\delta$-function, which is quite natural because the high temperature phase will eventually become free-fermion like.  

\begin{figure}[tp]
\includegraphics[width=3.0 in]
{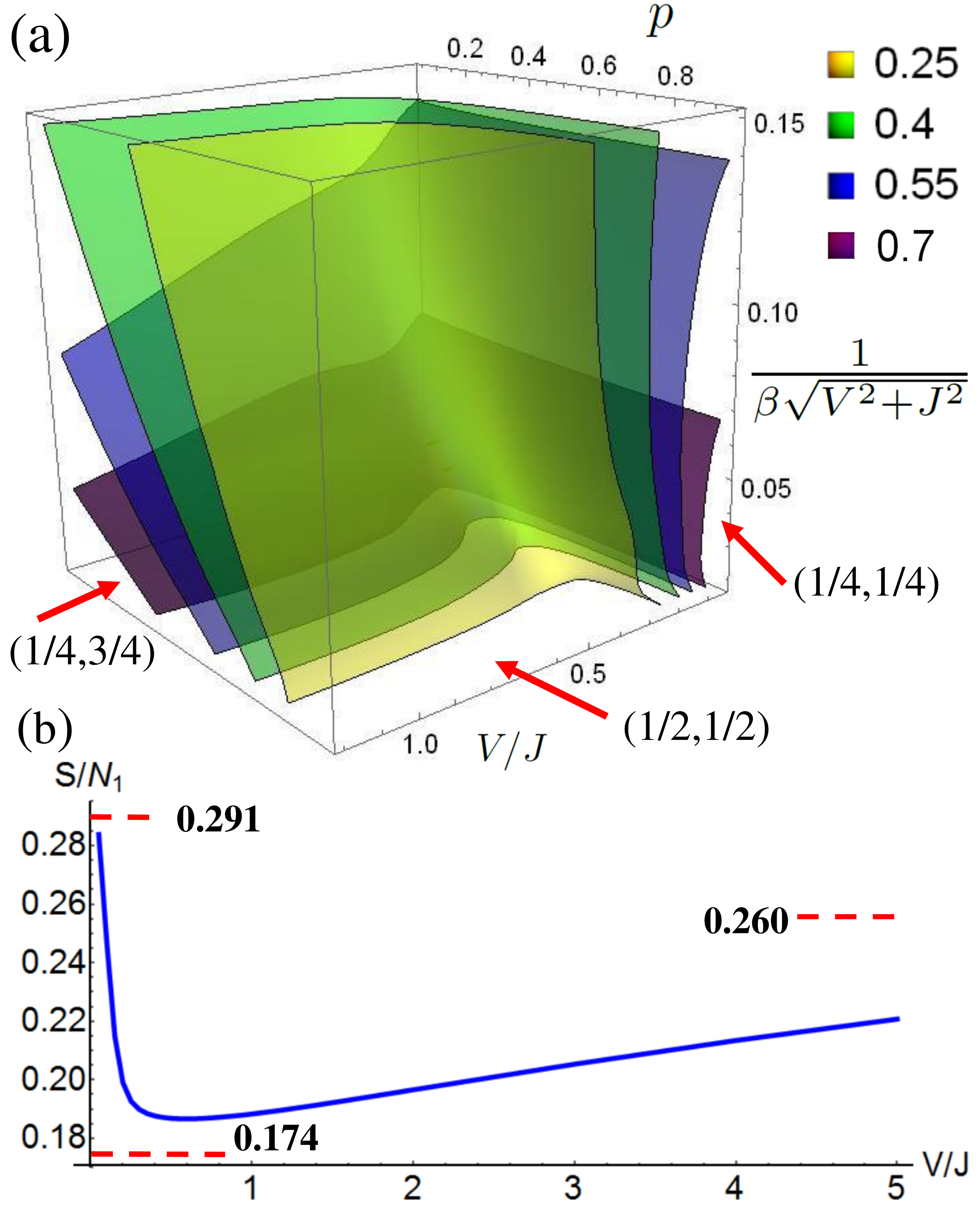}
\caption{(a) A contour plot of the Lyapunov exponent $\lambda/(2\pi k_\text{B}T)$ in a three-dimensional parameter space in term of $p$, $V/J$ and $1/(\beta\sqrt{J^2+V^2})$. The regions of different phases are indicated by the red arrows. (b). The entropy $S/N_1$ is plotted as a function of $V/J$, with $p$ fixed at $0.25$ and $1/(\beta\sqrt{J^2+V^2})$ fixed at $0.02$. The values marked by dashed lines are zero-temperature entropy for the $(1/4,1/4)$ phase, the $(1/4,3/4)$ phase and the $V/J\rightarrow\infty$ limit, respectively.}
\label{Lyapunov}
\end{figure}

As $V/J$ increases, the $(1/4,1/4)$ phase shrinks and the $(1/2,1/2)$ phase expands. Eventually, at large $V/J$, the $(1/4,1/4)$ phase gradually disappears and the $(1/2,1/2)$ phase directly connects to the high temperature free fermion phase. To view this more clearly, we draw a three-dimensional counter plot of $\lambda/(2\pi k_\text{B}T)$ in term of $p$, $V/J$ and $1/(\beta\sqrt{J^2+V^2})$, as shown in Fig. \ref{Lyapunov}. Without loss of generality, only $p<1$ regime is shown.  For a maximally chaotic phase, this value should approach unity at low-temperature. One can see from Fig. \ref{Lyapunov} that the regime nearby $V/J=0$ plane has a larger Lyapunov exponent, and this is the $(1/4,1/4)$ phase as it adiabatically connects to two decoupled SYK$_4$ in the $V/J=0$ plane. This regime indeed shrinks as $V/J$ increases. Another chaotic regime is the low-temperature regime with $p<1$, which is the $(1/4,3/4)$ phase. From the Lyapunov exponent, one can also see that, for fixed $V/J$, the closer $p$ is to unity, the lower temperature one needs in order to approach the upper bound for $\lambda$. This is consistent with Fig. \ref{spectral}(a) determined from the spectral function.   

In Fig. \ref{Lyapunov}(b), we show how the entropy changes as $V/J$ increase at low-temperature, with a fixed $p=0.25$. In this case, for small $V/J$, the entropy $S/N_1$ is very close to $(1+p)S_{SYK}=0.291$, as expected from the $(1/4,1/4)$ phase; and as $V/J$ increases, the entropy $S/N_1$ decreases toward $(1-p)S_{SYK}=0.174$, as expected from the $(1/4,3/4)$ phase. Further increasing $V/J$, $S/N_1$ actually gradually increases. This is because in the limit $V/J\rightarrow\infty$, the SYK$_2$ term dominates, which couples $N_1$ Majorana modes to $N_2$ Majorana modes and always leaves $N_1-N_2$ uncoupled modes. Hence, the entropy $S/N_1$ will eventually saturate to $|1-p|\frac{1}{2}\log 2=0.260$. This change of entropy is consistent with the phase diagram determined from the spectral function and the Lyapunov exponent. 

\textit{Outlook.} Our results illustrate interesting behaviors of quantum phase transitions between two non-Fermi liquid phases. Future works along this line can straightforwardly generalize our system from two SYK models to a SYK chain, with which one can investigate properties such as transport coefficients across the transition. Another aspect is that, since a single SYK model can also be understood from the gravity side by holographic duality, it will also be interesting to ask how to view the transition and the entire finite temperature phase diagram from the gravity side.   

\textit{Acknowledgement.} We thank Yu Chen for discussions. This work is supported by MOST under Grant No. 2016YFA0301600 and NSFC Grant No. 11325418 and Tsinghua University Initiative Scientific Research Program.

\textit{Note added.} Upon finishing this work, we became aware of a paper Ref.\cite{gen song} in which the authors studied a similar model that contains a chain of SYK models with same number of modes and coupled by the quadratic coupling. The focuses of these two works are different.

\end{document}